\documentclass[journal]{IEEEtran}
\usepackage[english]{babel}
\usepackage[utf8]{inputenc}
\usepackage[pdftex]{graphicx}
\usepackage{amsmath,mathrsfs}
\usepackage{algorithmic}
\usepackage{array}
\usepackage{mdwmath}
\usepackage{mdwtab}
\usepackage{eqparbox}
\usepackage[tight,footnotesize]{subfigure}
\usepackage[pdftex,usenames,dvipsnames]{color}
\usepackage{siunitx}
\usepackage{listings}
\usepackage[hidelinks]{hyperref}
\usepackage{parskip}
\usepackage{siunitx}
\usepackage{color}
\usepackage{comment}
\usepackage{breqn}
\usepackage{braket}
\definecolor{mygreen}{rgb}{0,0.6,0}
\definecolor{mygray}{rgb}{0.5,0.5,0.5}
\definecolor{mymauve}{rgb}{0.58,0,0.82}
\begin{document}

\title{On the Extension of Linear Damping to Quantum Mechanics through Fractionary Momentum Operators Pt.I}

\author{Luis Fernando Mora Mora\\\textit{luis.mora\_m@ucr.ac.cr}\\\textit{lmoramora94@gmail.com}\\~\IEEEmembership{Universidad de Costa Rica}}%

\markboth{}{}
\maketitle

\begin{abstract}
The use of fractional momentum operators and fractionary kinetic energy used to model linear damping in dissipative systems such as resistive circuits and a spring-mass ensambles was extended to a quantum mechanical formalism. Three important associated 1 dimensional problems were solved: the free particle case, the infinite potential well, and the harmonic potential. The wave equations generated reproduced the same type of 2-order ODE observed in classical dissipative systems, and produced quantized energy levels. In the infinite potential well, a zero-point energy emerges, which can be fitted to the rest energy  of the particle described by special relativity, given by relationship $E_r=mc^2$. In the harmonic potential, new fractional creation and destruction operators were introduced to solve the problem in the energy basis. The energy eigenvalues found are different to the ones reported by earlier approaches to the quantum damped oscillator problem reported by other authors. In this case, a direct relationship between the relativistic rest energy of the particle and the expected value of the fractionary kinetic energy in the base state was obtained. We conclude that there exists a relationship between fractional kinetic energy and special relativity energies, that remains unclear and needs further exploration, but also conclude that the current form of transforming fractionary momentum operators to the position basis will yield non-observable imaginary momentum quantities, and thus a correction to the way of transforming them needs to be explored further.   
\end{abstract}


\IEEEpeerreviewmaketitle


. 

\section{Introduction}

Fractional calculus is as old as its classical integer counterpart, but refers to integration and differentiation in non-integer orders. In the same way Euler's gamma function generalizes the concept of a factorial to any real or complex number, it allows the generalization of any integral or derivative to non-integer order \cite{li2015numerical,herrmann2011fractional,rudolf2000applications}. This branch of calculus has proven to model interesting phenomena such as electron collision in metals \cite{guia2016fractional}, quark confinement \cite{herrmann2011fractional}, has been related to fractals \cite{tatom1995relationship}, cosmological expansion \cite{rami2007differential}, and more importantly to this research, to non-conservative systems \cite{dreisigmeyer2003nonconservative,bauer1931dissipative,bateman1931dissipative,riewe1996nonconservative}. We will denote the fractionary derivative of order 1/2 on any variable \textbf{q} as the following, according to the Riemann-Louville definition of fractional  derivative:

\begin{equation}
    _aD_t^{1/2}[q] =  \frac{d}{dt}\Big[\int_a^t q(\tau)(t-\tau)^{-1/2} d\tau\Big] =q^{(1/2)} 
\end{equation}

Where the upper-index parenthesis on \textbf{q} indicate \textit{differentiation}, not to be confused with taking its square-root. It has been shown \cite{riewe1996nonconservative,bauer1931dissipative} that fractional order momentum operators can be used to formulate fractionary lagrangians of the form:

\begin{equation}
    \mathcal{L} = \frac{1}{2}m\dot{q}^2 + \frac{1}{2}Bq^{(1/2)^2} - V
\end{equation}

Which under the following fractionary Euler-Lagrange equations generate the equations of motion of dissipative systems \cite{riewe1996nonconservative,bauer1931dissipative,mora2020obtaining}.

\begin{equation}
    \frac{d}{dt}\Big(\frac{\partial \mathcal{L}}{\partial \dot{q}}\Big) + \frac{d^{(1/2)}}{dt^{(1/2)}}\Big(\frac{\partial \mathcal{L}}{\partial q^{(1/2)}} \Big) = \frac{\partial \mathcal{L}}{\partial q}
\end{equation}

The most known case of a dissipative system is the damped oscillator, which satisfies a homogenous differential equation of the form:

\begin{equation}
    m\Ddot{q} + B\dot{q} + m\omega^2q = 0  
\end{equation}

Where B is a positive constant, known as the ``viscous coefficient'' or the ``damping coefficient'' of the system, which is usually chosen fit to match impulse-response curves of the system being studied. This last equation is frequently reformulated in terms of a $\xi = B/2m$ factor in fields such as classical mechanics \cite{taylor2005classical,morin} and control engineering \cite{kuo1987automatic,ogata2002modern}:

\begin{equation}
    \Ddot{q} + 2\gamma\dot{q} + \omega^2q = 0
\end{equation}

When the coefficient $\gamma$ is positive, this type of differential equation has three types of solutions: 1) under-damped, 2) critically-damped and 3) over-damped solutions \cite{taylor2005classical,goldstein2002classical,morin}. As we shall shortly see, only the first case will be of interest to us in quantum mechanics. 

After studying how fractionary lagrangians provide the differential equations of linear dissipative systems modelled by the generalized ODE form of equation (4), the following question arises: How can this fractional formalism be translated to quantum mechanics? Let's start considering the Legendre transform of the lagrangian of equation (2), to obtain an homologous \textit{fractionary Hamiltonian operator}:

\begin{equation}
    \mathcal{H}_f = \frac{\textbf{P}^2}{2m} + \frac{\textbf{P}^{(1/2)^2}}{2B} + V
    \label{ec:fracham}
\end{equation}

Here, $\textbf{P}$ is the usual momentum operator, and the fractional momentum operator $\textbf{P}^{(1/2)}$ is the conjugate variable to the fractionary derivative of order 1/2 for the position coordinate, $q^{(1/2)}$. We shall call the fractionary order term in equation (6) the \textit{fractionary kinetic energy} of the particle, $\textbf{E}_f$. This fractionary momentum vector operator has been suggested to transform to the position basis as follows (see quation 9.11 in \cite{herrmann2011fractional}):

\begin{equation}
    P^{(1/2)}_i \xrightarrow{Position} -i(\hbar mc)^{1/2}\frac{d^{1/2}}{dx_i^{1/2}}
\end{equation}

The factor $(\hbar mc)^{1/2}$ ensures the correct momentum units for the fractional derivative of order 1/2 next to it, where c is the speed of light in the vacuum \cite{herrmann2011fractional}. In this way, if the fractional hamiltonian is applied in the time-independent eigenvalue energy problem of the form $H_f\ket{\psi} = E\ket{\psi}$, the following differential equation is obtained in the position basis:

\begin{equation}
    \nabla^2\psi + 2\xi\nabla^{(1/2)^2}\psi + \frac{2m}{\hbar^2}(E-V(r))\psi = 0
    \label{ec:dampedwave}
\end{equation}

Here, the factor $\xi = m^2c/\hbar B$ is taken to simplify notation, analogously to the $\gamma$ factor in equation (5). Also it should be added that the fractionary del operator $\nabla^{(1/2)}$ when squared is equivalent to the following scalar operator:

\begin{equation}
    \nabla^{(1/2)} = \hat{e}_i\frac{\partial^{1/2}}{\partial x_i^{1/2}} \implies \nabla^{(1/2)^2} = \frac{\partial}{\partial x}+\frac{\partial}{\partial y} + \frac{\partial}{\partial z}
\end{equation}

Then, equation 7 can be considered a modified version of the ``original'' Schr\"{o}dinger wave equation given by \cite{schroedinger1993quantization,griffiths2017introduction,shankar2012principles}:

\begin{equation}
    \nabla^2\psi + \frac{2m}{\hbar^2}(E-V(r))\psi = 0 
    \label{ec:wave}
\end{equation}

It should be said for the purpose of contrast that the original Schr\"{o}dinger's wave equation is generated by the hamiltonian operator:

\begin{equation}
    \mathcal{H} = \frac{\textbf{P}^2}{2m}+\textbf{V}
\end{equation}

\begin{equation}
    P_i \xrightarrow{Position} -i\hbar \frac{\partial}{\partial x_i}
\end{equation}
 
 Now we proceed to solve equation \ref{ec:dampedwave} for the free particle case in 1 dimension, for a box particle and lastly for the harmonic potential. 

\section{1 Dimensional Free Particle}

The first and simplest case to start studying the wave equation in expression (8) is the free particle case in 1 dimension. Taking V=0 in equation 7 results in the following 1 dimensional problem:

\begin{equation}
\frac{d^2\psi}{dx^2}+ 2\xi\frac{d\psi}{dx} + \frac{2mE}{\hbar^2}\psi = 0
\label{ec:qdo}
\end{equation}

Note that equation (13) now has the form of the generalized 1 dimensional damped oscillator as the one shown in equation (4). Equation (13) has the following characteristical polynomial:

\begin{equation}
\lambda^2 + 2\xi\lambda + \frac{2mE}{\hbar^2} = 0
\end{equation}

Thus the determinant which gives the solutions to be considered is:

\begin{equation}
    \Delta = 4\xi^2 - \frac{8mE}{\hbar^2}
\end{equation}

It was found that only the case where the determinant is negative will give solutions with quantized energies, because the positive solutions grow exponentially or linearly with distance and have no quantization conditions associated. The type of solution seeked is known as the ``under-damped'' solution mentioned earlier. This case corresponds to having $\Delta < 0$, which means:

\begin{equation}
    \xi^2 < \frac{2mE}{h^2}
    \label{ec:condition}
\end{equation}

The general solution to equation (13) is consequently, the following set of \textit{damped oscillations}:

\begin{equation}
    \psi(x) = e^{-\xi x}\Big(Ae^{\frac{i\sqrt{|\Delta|}x}{2}} + Be^{\frac{-i\sqrt{|\Delta|}x}{2}}\Big)
\end{equation}

Next if we take the wave number $k= \sqrt{|\Delta|}/2$ to simplify the notation, we can write the generalized solution as follows:

\begin{equation}
    \psi(x) = e^{-\xi x}\Big(Ae^{ikx} + Be^{-ikx}\Big)
    \label{ec:free}
\end{equation}

It is evident that this type of solution ``blows up'' for $x\xrightarrow{}-\infty$, which is an undesired effect for a probability distribution. This problem is very similar to the free particle problem for the original Schrödinger wave equation (10), in which sine and cosine functions appear as plane-wave solutions. Because sine and cosine functions are not normalizable for $x \in [-\infty,\infty]$, the plane-waves have to be taken as a superposition to accurately describe a quantum particle as a wave-packet. Following the same line of thought, for the solutions shown in equation (18) the following superposition of damped waves is proposed:

\begin{equation}
 \psi(x) = \int_{-\infty}^{\infty}A(k)e^{(-\xi+ik)x-\omega(k)t} dk 
\end{equation}
 
Here the temporal dependence of the waves has been introduced. This last equation is not quite a Fourier transform but it does become a Fourier transform when the coefficient $\xi = 0$. The amplitude A(k) contains the coefficients of the linear superposition of the damped oscillations solutions and can be obtained through an inverse transform when t=0.

In the case where $\xi$ is not a function of the wave number \textit{k}, the exponential damping can be taken out of the integral:

\begin{equation}
 \psi(x) = e^{-\xi x}\int_{-\infty}^{\infty}A(k)e^{i(kx-\omega(k)t)} dk = e^{-\xi x}u(x,t)
 \label{ec:packet}
\end{equation}

For example, lets consider a non-dispersive wave-packet moving to the right with velocity \textit{c}:

\begin{equation}
  u(x,t) =  e^{-(x-ct)^2+ik_0(x-ct)}
\end{equation}

With this type of wave-packet, equation (\ref{ec:packet}) reduces to:

\begin{equation}
    \psi(x) = e^{-\xi x}e^{-(x-ct)^2+ik_0(x-ct)}
\end{equation}

By completing squares this can be shown to be equivalent to a spatial traslation of the gaussian probability distribution:

\begin{equation}
    \psi(x) = A'(\xi,t)\cdot e^{-(x+\xi^2 -ct)^2 +ik_0(x-ct)}
\end{equation}

With A' equal to: 
\begin{equation}
A'=e^{-\xi ct + \xi^2/4}    
\end{equation}

So it appears that exponential damping can be applied on a gaussian wave packet,and it will only translate it. This corrects for the blowing-up effect of taking only one damped oscillation as $x\xrightarrow{}-\infty$. We will encounter again this type of translation of the gaussian envelope when solving the harmonic potential, but let's not get ahead of ourselves.

\section{Particle in a Box}

Now we set to solve equation (8) for the infinite potential well. We shall call this problem the ``damped'' potential well, in contrast with the ``non-damped'' problem that arises when solving the original Schrödinger equation with the same potential. If the box has a total length $L$ and we take the potential to be: 

\begin{equation}
  V(x) =
    \begin{cases}
      0 & \text{if x $\in$ [-L/2, L/2]}\\
      \infty & \text{otherwise}
    \end{cases}       
\end{equation}

The solution in the region where the potential is cero is given by equation (\ref{ec:free}). On the other hand, where the potential is infinite, the solution vanishes. The only thing left to do is to find the constants A and B in equation (18) for this problem using the boundary conditions $\psi(x = L/2) = 0$ and $\psi(x = -L/2) = 0$:

\begin{equation}
    e^{\xi L/2}(Ae^{\frac{-i\sqrt{|\Delta|}L}{4}} + Be^{\frac{-i\sqrt{|\Delta|}L}{4}}) = 0
\end{equation}

\begin{equation}
    e^{-\xi L/2}(Ae^{\frac{i\sqrt{|\Delta|}L}{4}} + Be^{\frac{i\sqrt{|\Delta|}L}{4}}) = 0
\end{equation}

These last equations can be re-written in matrix form as:

\begin{equation}
\begin{bmatrix} e^{-i\sqrt{|\Delta|}L/4} & e^{i\sqrt{|\Delta|}L/4} \\ e^{i\sqrt{|\Delta|}L/4} & e^{-i\sqrt{|\Delta|}L/4}   \end{bmatrix} \begin{bmatrix}A\\B\end{bmatrix}  = \begin{bmatrix}0\\0\end{bmatrix}  
\end{equation}

Forcing the determinant to be cero for non-trivial solutions in A and B, one gets the following condition:

\begin{equation}
e^{-i\sqrt{|\Delta|}L/2}-e^{i\sqrt{|\Delta|}L/2} = 0
\end{equation}

Multiplying and dividing on both sides by 2i, a energy quantization condition is obtained:

\begin{equation}
    2i \sin\Big(\frac{\sqrt{|\Delta|}L}{2}\Big) = 0
\end{equation}

\begin{equation}
\frac{\sqrt{|\Delta|}L}{2} = \pi n
\end{equation}

\begin{equation}
    |\Delta| = \Big(\frac{2\pi n}{L}\Big)^2
\end{equation}

Which means the allowed energy levels are:

\begin{equation}
    \frac{8mE}{\hbar^2}-4\xi^2 = \Big(\frac{2\pi n}{L}\Big)^2
\end{equation}

\begin{equation}
\implies E_n = \frac{m^3c^2}{2B^2} +  \frac{\hbar^2\pi^2n^2}{2mL^2}
\end{equation}

Thus the allowed energies are the same as in the non-damped solution \cite{shankar2012principles}, but a zero-point energy appears, that depend on the damping coefficient B. In fact, if the damping coefficient B is chosen as $B=m/\sqrt{2}$ then the zero-point energy of the particle coincides with the rest energy of the particle described by special relativity:

\begin{equation}
    B\xrightarrow[]{}\frac{m}{\sqrt{2}} \implies E_0 = mc^2 = E_r
\end{equation}

Though, for this n=0 case, the wave funtion vanishes, preserving the uncertainty principle (it could be interpreted as the probability of observing it at rest, which is zero if we know the energy of the particle). This allowed energies gives us the following wave numbers:

\begin{equation}
    k_n = \frac{2\pi n}{L}
\end{equation}

Note that the wave numbers obtained are even multiples of $\pi/L$ in contrast with the non-damped particle in a box \cite{shankar2012principles}. What happened to the odd wave numbers? If any of the two boundary conditions are reconsidered for the new wave numbers just found, one obtains that:

\begin{equation}
    A = -B 
\end{equation}

For this last condition we conclude the solutions for the ``damped'' particle in the box are only odd functions with even wave numbers, so that's why the odd wave numbers are left out. The solution then takes the general form:

\begin{equation}
\psi_{n}(x) = Ae^{-\xi x}\Big(e^{ik_nx} - e^{-ik_nx}\Big) = A_n e^{-\xi x}\sin(k_n x)    
\end{equation}

The $A'$ factor can be found through a normalization condition:

\begin{equation}
    \int_{-L/2}^{L/2} |\psi(x)_n|^2 dx = 1
\end{equation}

\begin{equation}
\implies A_n = \sqrt{2\xi\Big[\Big(\frac{\xi L}{2\pi n}\Big)^2 +1\Big]csch(\xi L)}
\end{equation}

Where csch(x) is the hyperbolic cosecant function. 

\section{1-Dimensional Quantum Mechanical Damped Oscillator}

Now we set to solve equation (8) for the harmonic potential in 1 dimension. We shall call this problem the ``damped quantum harmonic oscillator problem''. This problem was explored in early works \cite{kerner1958note,stevens1958wave} in which the authors used a exponential factor in the hamiltonian operator. We shall take a different approach by using the fractional hamiltonian formalism here developed to model the linear damping. Very different energy eigenvalues will be found using the fractional formalism, as we shall now show. Taking the potential as:

\begin{equation}
    V(x) = \frac{1}{2}m\omega^2x^2
\end{equation}

Makes equation (8) take the form:

\begin{equation}
\frac{d^2\psi}{dx^2}+ 2\xi\frac{d\psi}{dx} + \frac{2m}{\hbar^2}(E-\frac{1}{2}m\omega^2x^2)\psi = 0
\label{ec:qdo}
\end{equation}

In this case, an analogous procedure used by \cite{shankar2012principles} to solve the non-damped oscillator will be followed. The following change of variables is necessary to obtaining an a-dimensional scale:

\begin{equation}
    y = bx
\end{equation}

With $b=\sqrt{\hbar/m\omega}$. This leads the equation to take the form:

\begin{equation}
    \frac{d^2\psi}{dy^2} + \frac{m}{B}\sqrt{\epsilon_r}\frac{d\psi}{dy}+(2\epsilon -y^2)\psi=0
\end{equation}

Where $\epsilon = E/\hbar\omega$ and $\epsilon_r = mc^2/\hbar \omega$. Notice the equation is mostly equal to the non-damped quantum mechanical oscillator equation, except for the $d\psi/dy$ term that appears due to the fractionary kinetic energy operator introduced. By analyzing the two extreme cases when $y\xrightarrow{}0$ and $y\xrightarrow{}\infty$, the solution must have the form:

\begin{equation}
    \psi(y) = e^{-\frac{1}{2}(y^2+\frac{m}{2B}\sqrt{\epsilon_r}y)}M(y)
\end{equation}

Where M(y) is some sort of polynomial to be determined. To obtain a quantization condition, we wish to obtain a recursive relationship between the coefficients of these polynomial. To be able to find one, the solution form of equation (45) is re-inserted into equation (44) to obtain the following differential equation for M(y):

\begin{equation}
    \frac{d^2M}{dy} -2y\frac{dM}{dy} + (2\epsilon - \frac{m^2}{4B^2}\epsilon_r -1)M = 0
\end{equation}

Now, taking the polynomial as a power series of the form:

\begin{equation}
M(y) = \sum_{n=0}^\infty C_ny^m    
\end{equation}

And then re-inserting it into equation (46), the following recursive coefficient relationship is obtained:

\begin{equation}
    C_{n+2} = \frac{C_n(2n+1 + \frac{m^2}{4B^2}\epsilon_r -2\epsilon)}{(n+1)(n+2)}
\end{equation}

Which means, like in the case of the non-damped quantum mechanical oscillator, that there are either odd or even solutions, not both at the same time if we wish to have a normalizable solution. Luckily, the polynomials which satisfy this coefficient recursion are the familiar \textit{Hermite polynomials}. In order for the series to terminate, the numerator of equation (48) must equal zero, which finally gives the quantization condition for the $\epsilon$ energy levels:

\begin{equation}
    \epsilon_n = (n+\frac{1}{2}) +\frac{m^2}{8B^2}\epsilon_r
\end{equation}

\begin{equation}
    \implies E_n = \hbar\omega(n+\frac{1}{2}) + \frac{m^2}{8B^2}(mc^2)
\end{equation}

 Since the first Hermite polynomial is just a constant equal to 1, then the base-state for this damped oscillator is a \textit{displaced} gaussian distribution, like the one encountered in the free-particle case. Also it should be noted that at high frequencies, this solution coincides with the gaussian base state of the non-damped harmonic oscillator (because $\epsilon_r\xrightarrow[]{}0$): 
 
 \begin{equation}
    \psi_0(x) = A_0e^{-(\sqrt{\frac{\hbar}{m\omega}}x+\frac{m}{2B}\sqrt{\epsilon_r})^2} 
\end{equation}
 
 By normalizing the base state, the constant $A_0$ is found to be:
 
 \begin{equation}
    \int_{-\infty}^{\infty}|\psi_0(x)|dx = 1 \implies A_0 = \Big(\frac{m\omega}{\hbar\pi}\Big)^{1/4} 
 \end{equation}
 Which is the same for the non-damped harmonic oscillator. Also, it should be noticed how an additional term in the zero point energy appears compared with the zero point energy of $\hbar\omega/2$ in the non-damped harmonic oscillator. In fact, if we choose:

\begin{equation}
      B \xrightarrow[]{}\frac{m}{\sqrt{8}} \implies E_0 = mc^2 + (n+\frac{1}{2})\hbar\omega
\end{equation}

Then we can again recover the particle's relativistic rest energy of the particle $E_0 = mc^2$, plus the same energy increments in $\hbar\omega$ given by n ``quanta'' or particles, just like the non-damped harmonic oscillator \cite{shankar2012principles,griffiths2017introduction}. Under this value of B, the n-th level with energy eigenvalue $\hbar\omega (n+1/2) + mc^2$ wave function can be found to be of the form:

\begin{equation}
    \psi_n(y) = A_n H_n(y)e^{\frac{-1}{2}(y+\mu)^2}
\end{equation}

Where $H_n(y)$ is the n-th Hermite polynomial, $\mu = \sqrt{2\epsilon_r}$ and $A_n$ is a normalizing constant. The first 3 normalizing constants have the form: 

\begin{equation}
    A_1 = \frac{1}{\sqrt{1!+2^1\mu^2}}\Big(\frac{m\omega}{\hbar\pi 2^2}\Big)^{1/4}
\end{equation}

\begin{equation}
    A_2 = \frac{1}{\sqrt{2!+2^2\cdot2\mu^2+2^2\mu^4}}\Big(\frac{m\omega}{\hbar\pi 2^4}\Big)^{1/4}
\end{equation}

\begin{equation}
    A_3 = \frac{1}{\sqrt{3!+\frac{3^2}{2}8\mu^2+\frac{3^2}{1}4\mu^4 +2^3\mu^6}}\Big(\frac{m\omega}{\hbar\pi 2^6}\Big)^{1/4}
\end{equation}

Such that the n-th normalizing constant can be generalized to be of the form:

\begin{equation}
    A_n = \frac{1}{\sqrt{P_n(\mu)}}\Big(\frac{m\omega}{\hbar\pi 2^{2n}}\Big)^{1/4}
\end{equation}

Where $P_n(\mu)$ is a polynomial in even powers of $\mu$, with coefficients $G_{n,m}$ that can be calculated recursively from the previous $P_{n-1}(\mu)$ polynomial's coefficients $G_{n-1,m}$:

\begin{equation*}
    P_n(\mu) = \sum_{k=0}^n G_{n,2k}\cdot\mu^{2k} = n! + 2^{n}\mu^{2n}+n^2\sum_{k=1}^{n-1} \frac{G_{n-1,2k}}{n-k}\mu^{2k} 
\end{equation*}

It should be noticed that when $\mu=0$, then the polynomials $P_n(\mu)$ reduce to only $n!$ and in this case the normalizing constants $A_n$ agree with the ones in the non-damped harmonic oscillator. 

\section{Quantum Mechanical Damped Oscillator in the Energy Basis}

To solve the same problem we just solved but now in the energy basis with energy eigenkets $\ket{\epsilon_n}=\ket{n}$ we now introduce a modified version of the destruction and creation operators. We shall call them ``fractional'' because they produce eigenstates for the fractional hamiltonian operator, even though there are no fractional derivatives involved in their formulation. These fractional destruction and creation are constructed respectively as:

\begin{equation}
\textbf{a}_f = \sqrt{\frac{m\omega}{2\hbar}}\textbf{X} + \frac{m}{2B}\sqrt{\frac{\epsilon_r}{2}} + i\sqrt{\frac{1}{2\hbar m\omega}}\textbf{P} 
\end{equation}

\begin{equation}
\textbf{a}^\dagger_f = \sqrt{\frac{m\omega}{2\hbar}}\textbf{X} - \frac{m}{2B}\sqrt{\frac{\epsilon_r}{2}} - i\sqrt{\frac{1}{2\hbar m\omega}}\textbf{P} 
\end{equation}

Using the following commutator property (which can be easily shown but its proof will be ommited):

\begin{equation}
    [\textbf{a}_f,\textbf{P}^{(1/2)^2}] = \hbar mc\sqrt{\frac{m\omega}{2\hbar}}
\end{equation}

It can be shown that the fractional creation and destruction operators follow the desired properties:

\begin{equation}
    \mathcal{H}_f \textbf{a}_f\ket{\epsilon} = (\epsilon-1)\textbf{a}_f\ket{\epsilon}
\end{equation}

\begin{equation}
    \mathcal{H}_f \textbf{a}^\dagger_f\ket{\epsilon} = (\epsilon+1)\textbf{a}_f\ket{\epsilon}  
\end{equation}

Which means $\textbf{a}_f\ket{\epsilon}$ is an eigenstate of the fractional hamiltonian $\mathcal{H}_f$ with energy eigenvalue $\epsilon-1$, and analogously $\textbf{a}^\dagger_f\ket{\epsilon}$ is also an eigenstate with energy eigenvalue $\epsilon+1$. Since we know that the energy eigenvalues are positive, then there must exists a ground state unto which the destruction operation destroys the state:

\begin{equation}
    \textbf{a}_f\ket{\epsilon_0} = 0
\end{equation}

\begin{equation}
    \implies \Big[\sqrt{\frac{m\omega}{2\hbar}}x + \frac{m}{2B}\sqrt{\frac{\epsilon_r}{2}} -\hbar\sqrt{\frac{1}{2\hbar m\omega}}\frac{d}{dx}\Big]\psi_0(x) = 0
\end{equation}

And by separating variables and integrating the last equation, similarly as \cite{shankar2012principles} $\psi_0(x)$ is found to be:

\begin{equation}
    \psi_0(x) = A_0\cdot e^{-(\sqrt{\frac{\hbar}{m\omega}}x+\frac{m}{2B}\sqrt{\epsilon_r})^2} 
\end{equation}
 
Which matches the solution found earlier in the position basis. This allows the solutions to be written in the form:

\begin{equation*}
    \psi_n(x) = \frac{1}{\sqrt{P_n(\mu)}}\Big[\sqrt{\frac{m\omega}{2\hbar}}x + \frac{m}{2B}\sqrt{\frac{\epsilon_r}{2}} -\hbar\sqrt{\frac{1}{2\hbar m\omega}}\frac{d}{dx}\Big]^n\psi_0(x)
\end{equation*}

In the light of these results and by comparing the generalized form of the normalizing constants shown in equation (58), then it is evident that the fractional destruction operator acts like:

\begin{equation}
    \textbf{a}_f\ket{n} = \sqrt{\frac{P_{n}(\mu)}{P_{n-1}(\mu)}}\ket{n-1}\xrightarrow[]{\mu=0} \sqrt{n}\ket{n-1}
\end{equation}

And analogously the fractional creation opeator acts like:

\begin{equation}
    \textbf{a}^\dagger_f\ket{n} = \sqrt{\frac{P_{n+1}(\mu)}{P_{n}(\mu)}}\ket{n+1} \xrightarrow[]{\mu=0}\sqrt{n+1}\ket{n+1}
\end{equation}

 Which leads us to conclude that the fractional creation and destruction operators are a generalization of the non-damped oscillator ladder operators, which return the same results when $\mu=0$. This way, arbitrary eigenstates can be expressed in terms of the base-state $\ket{0}$ by succesively applying the fractional creation operator:
 
 \begin{equation}
     \ket{n} = \frac{(\textbf{a}^\dagger_f)^n}{\sqrt{P_n(\mu)}}\ket{0} \xrightarrow[]{\mu = 0} \frac{(\textbf{a}^\dagger)^n}{\sqrt{n!}}\ket{0}
 \end{equation}

Which again, return the non-damped oscillator relationship for arbitrary eigenstates when $\mu=0$. Thus, a general state of composed of several damped harmonic oscillators can be built by acting on the vacuum state with the corresponding fractional creation operators:

\begin{equation}
     \ket{n_1n_2\dots} = \prod_k \frac{(\textbf{a}^\dagger_{f,k})^{n_k}}{\sqrt{P_{n_k}(\mu)}}\ket{0} 
 \end{equation}

\subsection{Quantization of the Fractionary Energy in the Damped Harmonic Oscillator}

Using the fractional creation and destruction operators, the fractional hamiltonian can be re-written as:

\begin{equation*}
    \mathcal{H}_f = \hbar\omega\Big[(\textbf{a}^\dagger_f +\frac{m}{2B}\sqrt{\frac{\epsilon_r}{2}})(\textbf{a}_f -\frac{m}{2B}\sqrt{\frac{\epsilon_r}{2}})+\frac{1}{2}\Big]+ \frac{\textbf{P}^{(1/2)^2}}{2B}
\end{equation*}

And taking the expected value on the n-th energy level on both sides of the equation then implies that:

\begin{equation*}
\hbar\omega n + \frac{m^2}{8B^2}(mc^2) = \hbar\omega \frac{P_n(\mu)}{P_{n-1}(\mu)} - \frac{m^2}{8B^2}(mc^2) + \Big<\frac{\textbf{P}^{(1/2)^2}}{2B}\Big>
\end{equation*}

Which means we just found that the fractionary kinetic energy in this problem is also quantized:

\begin{equation}
    \implies \Big<\frac{\textbf{P}^{(1/2)^2}}{2B}\Big> = \hbar\omega\Big(n- \frac{P_n(\mu)}{P_{n-1}(\mu)}\Big)+\frac{m^2}{4B^2}(mc^2)
\end{equation}

By the condition $B = m/\sqrt{8}$ that we chose earlier, this fractionary energy reduces to:

\begin{equation}
     \Big<\textbf{E}_f\Big> = \hbar\omega\Big(n-\frac{P_n(\mu)}{P_{n-1}(\mu)}\Big)+2(mc^2)
\end{equation}

This relationship can also be inverted to put the relativistic rest energy of the particle in terms of the expected value of fractionary kinetic energy:  

\begin{equation}
     E_r = mc^2 = \frac{1}{2}\Big<\textbf{E}_f\Big> - \frac{\hbar\omega}{2}\Big(n-\frac{P_n(\mu)}{P_{n-1}(\mu)}\Big)
\end{equation}

For the base state where n=0, the polynomial fraction vanishes because the $\textbf{a}_f$ destroys the $\ket{0}$ state and correspondingly the expected value. Evaluating n=0 for equation (73) then shows a direct relationship between the relativistic rest energy of the particle and the expected value of the fractionary kinetic energy in the base state:

\begin{equation}
    mc^2 = \frac{1}{2}\Big<\textbf{E}_f\Big>_{n=0}
\end{equation}

\section{Expected Momentum Values}

The theory developed until now has shown no inconsistencies. But now the expected values of momentum $\textbf{P}$ will be considered and shown to be imaginary, making the transformation in equation (7) suggested by \cite{herrmann2011fractional} non suitable for observable physics. 

\subsection{Expected Value of Momentum of the Particle in the Infinite Well Potential}

Taking the expected value of the momentum operator with the solutions found in expression (38):

\begin{equation*}
    \bra{n}\textbf{P}\ket{n} = \int_{-\infty}^{\infty} \psi(x)_n (-i\hbar)\frac{d}{dx}\psi(x)_n
\end{equation*}

\begin{equation*}
     = -i\hbar A_n^2\int_{-L/2}^{L/2} e^{-2\xi x}sin(k_nx)[-\xi sin(k_nx)+k_ncos(k_nx)]
\end{equation*}

And since the normalization constant is real as same as the integral, then the expected value of momentum is an imaginary quantity and thus non-observable. 

\subsection{Expected Value of the Particle in the Harmonic Oscillator}

Expressing the momentum operator as a combination of the fractional creation and destruction operators:

\begin{equation}
    \bra{n}\textbf{P}\ket{n} = \bra{n}\frac{\sqrt{2\hbar m\omega}}{2i}(\textbf{a}_f - \textbf{a}_f^\dagger -\frac{m}{B}\sqrt{\frac{\epsilon_r}{2}})\ket{n} 
\end{equation}
And by the orthonormality properties of the solutions found then:
\begin{equation}
    \bra{n}\textbf{P}\ket{n} = \frac{\sqrt{2\hbar m \omega \epsilon_r}}{2i\sqrt2}
\end{equation}

Which is another imaginary and unobservable quantity.

\section{Conclusions}

The linear damping effect considered in dissipative systems, modelled through the means of fractionary momentum operators, was succesfully extended to a quantum mechanical formalism. A modified wave-equation was obtained and solved first for the free particle case, showing that a wavepacket-like solution is possible as the damping coefficient later chosen is constant and only translates gaussian packets. 

The modified wave equation was solved for the infinite potential well case, and was shown to have discrete energy levels with a zero point energy, admitting only odd damped-oscillations as solutions. This zero point energy matches the relativistic rest energy of the particle when choosing a damping coefficient equal to $m/\sqrt{2}$. 

Thirdly, the same wave equation was solved for the harmonic potential, and an additional energy term was found in the eigenvalues, compared to the non-damped harmonic oscillator. This result is different and more useful compared to the ones reported in earlier works by other authors. The zero point energy in this case can be also be matched to the relativistic rest energy of the particle when choosing a damping coefficient of $B=m/\sqrt{8}$. From this result, a direct relationship between the expected value of the fractionary kinetic energy of the particle and the relativistic rest energy was shown.

This results show that there might exist an unexplored relationship between fractionary energy operators of order 1/2 and transformations which arise in special relativity, which remains unclear. However, the transformation of fractionary momentum operators suggested by the literature does not yield a theory compatible with the classical limit, which describes the dissipation of energy through fractionary energies. In a future article, a different transformation for the fractionary momentum operator will be considered and will be associated with real valued momentum observables.  

\bibliographystyle{IEEEtran}
\bibliography{bibliography}

\end{document}